\documentclass[12pt]{article}
\begin{document}
\title{ Realization of the spectrum generating algebra for the generalized Kratzer potentials.\footnote{ {\bf Published} in International Journal of Theoretical Physics 2010 {\bf 49},1303.     }   }

\author{ \bf {OYEWUMI, K. J.}  \footnote{ E-Mail: mjpysics@yahoo.com~}\\
Theoretical Physics Section, Department of Physics, \\University of Ilorin, P. M. B. 1515, Ilorin, Kwara state, Nigeria.\\
and\\
International Chair of Mathematical Physics and Applications\\ (ICMPA-UNESCO Chair)\\ Universite d'Abomey-Calavi, Cotonou, Republic of Benin. }

\maketitle

\begin{abstract}
The dynamical symmetries of the Kratzer-type molecular potentials (generalized Kratzer molecular potentials) are studied by using the factorization method. The creation and annihilation (ladder) operators for the radial eigenfunctions satisfying quantum dynamical algebra $SU(1, 1)$ are established. Factorization method is a very simple method of calculating the matrix elements from these ladder operators. The matrix elements of different functions of $r$, $r\frac{d}{dr}$, their sum $\Gamma_{1}$ and difference $\Gamma_{2}$ are evaluated in a closed form. The exact bound state energy eigenvalues $E_{n, \ell}$ and matrix elements of $r$, $r\frac{d}{dr}$, their sum $\Gamma_{1}$ and difference $\Gamma_{2}$  are calculated for various values of $n$ and $\ell$ quantum numbers for $CO$ and $NO$ diatomic molecules for the two potentials. The results obtained are in very good agreement with those obtained by other methods.
\end{abstract}


\baselineskip=0.28in \vspace{2mm}
 \vspace{2mm}

\section{Introduction}
Exactly solvable models in Physics have been generating a renewed
interest, because of the fact that they can be solved in terms of
creation and annihilation operators by means of factorization method
[1 - 5]. In the year 2002, this present method was proposed in the
two related works: the ladder operators for the modified
P\"{o}schl-Teller potential and the  Morse potential were obtained
\cite{DoE02DoL02}.  With factorization method, the ladder operators
of a quantum mechanical system with some important potentials like
Morse potential, P\"{o}schl-Teller potential, the pseudoharmonic
potential, the infinitely square-well potential and other quantum
systems have been established \cite{DoE02DoL02,DEE}. For
comprehensive review (see Dong and references therein \cite{Don07}).

Algebraic methods underlying Lie symmetry and its associated algebra
have been widely used to study many of these exactly solvable
potentials. In 2007, Rasinariu et al. \cite{RaE07} gave a review of
the progress made so far in solving exactly solvable problems in
quantum mechanics, by connecting supersymmetry and spectrum
generating algebras through the property of shape invariance. The
solutions of exactly solvable models can be achieved by dynamical
algebraic approaches [1 - 11]. In 1991, De Lange and Raab
\cite{DeR91} presented  operator methods with shift operators (that
is, raising and lowering operators) for the Hamiltonian of exactly
solvable models.

For the factorization method, we are going to adopt Dong's approach \cite{Don07} by finding the ladder operators $K^{\pm}$ with the following properties:
\begin{equation}
K_{n}^{\pm} \Phi_{n}(x) = k_{n}\Phi_{n + 1}(x).
\label{m1}
\end{equation}
In this case, we  are seeking for the ladder operators of the form
\begin{equation}
K_{n}^{\pm}  = A_{n}^{\pm}(x)\frac{d}{dx} + B_{n}^{\pm}(x),
\label{m2}
\end{equation}
these ladder operators depend on the physical variable $x$, which is different for different  quantum systems. With equation(\ref{m1}), the following expressions are obtained:
\begin{equation}
K_{n + 1}^{-} K_{n}^{+}\Phi_{n}(x) = k_{n}^{+} K_{n + 1}^{-}\Phi_{n}(x),~~~K_{n}^{+} K_{n + 1}^{-}\Phi_{n + 1}(x) = k_{n}^{+} K_{n + 1}^{-}\Phi_{n + 1}(x),
\label{m3}
\end{equation}
from which we find that the products of the operators $K_{n + 1}^{-} K_{n}^{+}$ and $K_{n + 1}^{+} K_{n}^{-}$ acting on the given wavefunctions $\Phi_{n}(x)$ and $\Phi_{n + 1}(x)$, respectively, have the same constant
\begin{equation}
c_{n} = k_{n}^{+} K_{n + 1}^{-}.
\label{m4}
\end{equation}

 The $SU(1, 1)$ algebra has useful applications in Physics \cite{Don07,Ada94,OyE08,OyK08}. The algebra of the group $SU(1, 1)$ is used to generate the energy spectra while the representation matrices of the group could be used to calculate time dependent excitations of the bound states and the scattering states respectively.

For diatomic molecules, the generalised Kratzer potentials is
considered as one of the molecular potentials. These potentials have
the general features of the true interaction
 energy, interatomic, inter-molecular and dynamical properties, their wavefunctions do
 vanish at the origin like Morse potential. This added advantage make these potentials
 important  in Molecular Physics, Chemical Physics, Solid State Physics. These potentials (Kratzer and modified Kratzer molecular
 potentials)~that is being considered in the present work are important molecular potentials
 which describe the interaction between two atoms .
 These potentials have  been used extensively to describe the molecular structure and
  interactions and have been receiving much attention in the history of quantum chemistry
  for some decades ago \cite{ BaE07,Flu94, BeE06, IkS08, Oye05, KrF2026}.

It is the purpose of this paper to study the dynamical symmetries of the Kratzer-type molecular potentials (generalized Kratzer molecular potentials) and to establish that the generators obtained are an $SU(1, 1)$  dynamical group,  and to obtain some numerical results for the energy eigenvalues and matrix elements for some diatomic molecules.

The paper is organized as follows. In Section $2$, we study the
exact solutions of the Kratzer-type molecular potentials. Section
$3$ contained the construction of the creation and annihilation
operators, the commutation relations of some of these operators and
the matrix elements of some related functions $r$ and
$r\frac{d}{dr}$. Also, in Section $4$, the numerical calculations of
the energy levels and matrix elements for some diatomic molecules
for the Kratzer and modified Kratzer potentials are given. We
conclude with Section $5$.

\section{Exact solutions of the Kratzer-type molecular potentials}
Consider the molecular potential (the generalized Kratzer potential)
of the form \cite{OyK08,IkS071,IkS072,IkS08}
\begin{equation}
V(r) =\frac{a}{r} + \frac{b}{r^{2}} + c.
\label{m5}
\end{equation}
This potential is of great interest because, it is a generalised form of the
molecular potential which can be used to generate other forms of the Kratzer-type
molecular potentials:

\begin{itemize}
    \item Standard Morse or Kratzer-Fues potential
    \begin{equation}
V(r) = -D_{0}\left(\frac{2 r_{0}}{r} - \frac{r_{0}^{2}}{r^{2}}\right)
\label{m6}
\end{equation}
\end{itemize}
where $D_{0}$ is the dissociation energy between two atoms in a solid and
 $r_{0}$ is the equilibrium intermolecular separation. On comparing with equation (5),
 we have $a = -2D_{0}r_{0}, b = D_{0}r_{0}^{2}, c = 0$ and this potential has minimum
 value to be $-D_{0}$ at $r = r_{0}$ [2, 12 - 14, 17 - 23]. 
\begin{itemize}
\item Modified Kratzer potential
\begin{equation}
V(r) = -D_{0}\left(\frac{2 r_{0}}{r} - \frac{r_{0}^{2}}{r^{2}}\right) + D_{0}~=~D_{0}\left(\frac{r -r_{0}}{r} \right)^{2},
\label{m7}
\end{equation}
\end{itemize}
where, $a = -2 D_{0} r_{0}$, $b =  D_{0} r_{0}^{2}$ and $c =  D_{0} $.~
Modified  Kratzer potential is Kratzer potential shifted by the
amount $D_{0}$ \cite{BeE06,IkS071,IkS072,IkS08}.
The graphs of the two potentials (equations $6$ and $7$) are shown below in {\bf Figure 1.}

\noindent
{\bf Figure 1.} Shapes of the Kratzer and modified Kratzer
potentials for $CO$ and $NO$ diatomic molecules.

Consider the motion of a particle in a spherically symmetric potential $V(r)$
\begin{equation}
- \frac{\hbar^{2}}{2 \mu}\Delta \Psi(r,\theta, \phi) = [E - V(r)]\Psi(r,\theta, \phi)
\label{m8}
\end{equation}
where the potential $V(r)$ is taken as in equation (\ref{m5}).
We seek for the wavefunctions of the form
\begin{equation}
\Psi_{n, \ell, m}(r,\theta, \phi)=R_{n, \ell}(r)Y_{\ell}^{m}(\theta, \phi)
\label{m9}
\end{equation}
which reduces equation (\ref{m8}) into the radial and angular wave functions as:
\begin{equation}
\frac{d~^{2} R_{n, \ell}~(r) }{dr~^{2}} + \frac{2}{r} \frac{d R_{n, \ell}~(r)}{dr} + \left\{ \frac{2 \mu}{\hbar^{2}} \left[E - \left(\frac{a}{r} + \frac{b}{r^{2}} + c \right)  \right] - \frac{\ell~(\ell + 1)}{r^{2}}   \right\}R_{n, \ell}~(r)= 0
\label{m10}
\end{equation}
and
\begin{equation}
L^{2}Y_{\ell }^{m}(\theta, \phi) - \hbar^{2} \ell~(\ell + 1)Y_{\ell}^{m}(\theta, \phi) = 0,
\label{m11}
\end{equation}
where $R_{n, \ell}(r)$ and $Y_{\ell }^{m}(\theta, \phi)$ are the radial and angular solutions of equations (\ref{m10} and \ref{m11}) respectively. For $\Psi_{n, \ell, m}~(r, \theta, \phi)$ to be finite everywhere, $R_{n,~\ell}~(r)$ must vanish at $r = 0$, that is, $R_{n,\ell}~(0) = 0$, then $R_{n,~\ell}~(r)$ is a real function.

For the bound state energy eigenvalues for this quantum system, the following dimensionless abbreviations are introduced :

\noindent
\begin{equation}
\rho = \gamma^{\frac{1}{2}}r = \left[ \frac{-8 \mu}{\hbar^{2}} (E - c) \right]^{\frac{1}{2}}r;~~\alpha = \left( \frac{- \mu}{2 \hbar^{2}(E - c)} \right)^{\frac{1}{2}}a;~~\beta_{\ell}~(\beta_{\ell} + 1) = \frac{2 \mu b}{\hbar^{2}} + ~ \ell~(\ell + 1)
\label{m12}.
\end{equation}
The above equations (\ref{m12}a) and (\ref{m12}b) are chosen in order that the acceptable bounds state solutions are obtained, this can only be possible if $E<0$ (otherwise, continuum state solutions will be obtained). $\beta_{\ell}$ in equation (\ref{m12}c) represents the usual centrifugal term, that is, $\beta_{\ell}~(\beta_{\ell} + 1) = \frac{2 \mu b}{\hbar^{2}} + ~ \ell~(\ell + 1)$, where $\ell$ is the angular momentum quantum number. These substitutions allow us to obtain the following hypergeometric-type equation.
\begin{equation}
\frac{d^{2}}{d \rho^{2}}R_{n, \ell}~(\rho) + \frac{2}{\rho} \frac{d }{d \rho } R_{n, \ell}~(\rho) + \left[- \frac{1}{4} + \frac{\alpha}{\rho} - \frac{\beta_{\ell}(\beta_{\ell} + 1)}{\rho^{2}}  \right]R_{n, \ell}~(\rho) = 0.
\label{m13}
\end{equation}

This differential equation has an irregular singularity as $\rho \rightarrow \infty$, where its normalized solutions in bound states behave like $exp~(\rho)$. It further has a singularity at $\rho \rightarrow 0$, where $R_{n,~\ell}~(\rho) \sim \rho^{\beta_{\ell}}$. Then, the ansatz for the wave functions which is a physically acceptable solution for $R_{n,~\ell}~(\rho)$ can be expressed in the form
\begin{equation}
R_{n, \ell}~(\rho)  = N_{n,\ell}~e^{- \rho/2} \rho^{\beta_{\ell}}G(\rho)
\label{m14}
\end{equation}
and therefore, equation (\ref{m13}) becomes
\begin{equation}
\rho \frac{d^{2}G(\rho)}{d\rho^{2}} + [(2 \beta_{\ell} + 2)- \rho   ] \frac{d G(\rho) }{d \rho} - [\beta_{\ell}  + 1 - \alpha ]G( \rho) = 0.
\label{m15}
\end{equation}
This is the associated Laguerre differential equation (Kummer equation)\cite{Flu94,GrR94}, the solution of equation (\ref{m15}) which is regular at origin (regular at $r = 0$ or $\rho = 0$) is the degenerate hypergeometric function
\begin{equation}
G(\rho) = ~_{1}F_{1}(\beta_{\ell}  + 1 - \alpha ,~  2 \beta_{\ell}  + 2 ; ~\rho).
\label{m16}
\end{equation}
For large values of $\rho$, this solution diverges as $exp(\rho)$, thus preventing normalization, except for when
\begin{equation}
\beta_{\ell} + 1 - \alpha = -n_{r} = -n; ~~n_{r} = 1,  2, \ldots
\label{m17}
\end{equation}
becomes a polynomial.

Therefore, the solution for the radial equation for this generalized Kratzer-type molecular potential is
\begin{equation}
R_{n, \ell}~(\rho)  = N_{n,\ell}~e^{- \rho /2} \rho^{\beta_{\ell}}
~~_{1}F_{1}(-n ,~  2 \beta_{\ell}  + 2 ; ~\rho),
\label{m18}
\end{equation}
where the normalization $N_{n_{r},\ell}$ is determined from the requirement that
\begin{equation}
\int_{0}^{\infty} \mid R_{n, \ell}~(\rho) \mid^{2}r^{2}dr  = 1.
\label{m19}
\end{equation}

By using the expression that relates the associated Laguerre functions with the confluent hypergeometric functions
\begin{equation}
_{1}F_{1}(-\gamma,~  m + 1; ~z) = \frac{\gamma~! m~!}{(\gamma + m)~!} ~L_{\gamma}^{m}(z)
\label{m20}
\end{equation}
together with the following important formular \cite{Flu94, GrR94}
\begin{equation}
\int_{0}^{\infty} e^{-x} x^{a} L_{n}^{a - 1}(x)L_{m}^{a - 1}(x) dx = \frac{(a + 2n) \Gamma(a + n) }{n~!}\delta_{nm}
\label{m21},
\end{equation}
and by substituting $\gamma = \xi^{2} $, the normalized radial wave function is obtained as
\begin{equation}
R_{n, \ell}~(r)  = N_{n,\ell}~e^{- \frac{\xi r}{2}} r^{ \beta_{\ell} }L_{n}^{2  \beta_{\ell} + 1}(\xi r),
\label{m22}
\end{equation}
where
\begin{equation}
N_{n, \ell}  = \left[ \frac{\xi^{2 \beta_{\ell} + 3}}{2} \frac{n!}{(n + \beta_{\ell} + 1)\Gamma(n + 2\beta_{\ell} + 2)}  \right]^{\frac{1}{2}}.
\label{m23}
\end{equation}

The corresponding eigenvalues are :
\begin{equation}
E_{n, \ell}  =  \frac{- \mu a^{2}}{ 2 \hbar^{2}  \left[ n + \beta_{\ell} + 1 \right]^{2}} + c,
\label{m24}
\end{equation}
where $\beta_{\ell}$ is obtained as the positive root of equation (\ref{m12}c) given as
\begin{equation}
\beta_{\ell}  =  \frac{1}{2} \left[-1 + \sqrt{( 2 \ell + 1)^{2}  + \frac{8 \mu b}{\hbar^{2}}}~~  \right].
\label{m25}
\end{equation}

\section{Construction of the creation and annihilation operators}
The ladder operators can be generated directly from the eigenfunction with the factorization method as shown in Dong \cite{Don07} and references therein. We shall find the diferential operators $\hat{L}_{\pm}$ with the following property:
\begin{equation}
\hat{L}_{\pm}R_{n,\ell}~(r)=\ell_{\pm}R_{{n \pm}1,\ell}~(r),
\label{m26}
\end{equation}
the operators of the form
\begin{equation}
\hat{L}_{\pm}=A_{\pm}(r)\frac{d}{dr} + B_{\pm}(r)
\label{m27}
\end{equation}
which depend only on the physical variable $r$ are to be obtained.

The action of the differential operator $\frac{d}{dr}$ on wave functions (\ref{m22}) gives:
\begin{equation}
\frac{d }{dr} R_{n,\ell}~(r) = -\frac{\xi}{2} R_{n,\ell}~(r)  +  \frac{\beta_{\ell}}{r} R_{n,\ell}~(r) +  N_{n, \ell}~ r^{\beta_{\ell}}~e^{-\frac {\xi r}{2}} \frac{d}{dr}L_{n}^{2 \beta_{ \ell} + 1}( \xi r ).
\label{m28}
\end{equation}
The expression above is used to construct the ladder operators $\hat{L}_{\pm}$ by using the recurrence relations of the associated Laguerre functions in order to find the relation between $R_{n,\ell}~(r)$ and $R_{n + 1,\ell}~(r)$. To find these, the following recurrence relations of the associated Laguerre functions are used \cite{Flu94,GrR94}:
\begin{equation}
x \frac{d}{dx}L_{n}^{\alpha}(x) = \left\{
\begin{array}{ll}
n L_{n}^{\alpha}(x) - (n + \alpha)L_{n-1}^{\alpha}(x) & \\ 
(n + 1) L_{n + 1}^{\alpha}(x) - (n + \alpha + 1 - x)L_{n}^{\alpha}(x)
 & \\ 
\end{array}
\right.
\label{m29}
\end{equation}
and the creation and annihilation operators are obtained as:
\begin{equation}
\hat{\cal{L}}_{-}  =  - r \frac{d }{dr}  - \frac{\xi r}{2} + \hat{n} + \beta_{\ell}~, ~~~~\hat{\cal{L}}_{+}  = r \frac{d }{dr}  - \frac{\xi r}{2} + \hat{n} + \beta_{\ell}~ + 2,
\label{m30}
\end{equation}
where $\hat{n}$ is the number operator with the property
\begin{equation}
\hat{n}R_{n,\ell}~(r) = nR_{n,\ell}~(r).
\label{m31}
\end{equation}

The action of the creation and annihilation operators on the radial wavefunctions $R_{n, \ell}~(r)$ gives the following properties
\begin{equation}
\hat{\cal{L}}_{\pm}R_{n,\ell}~(r)  =\ell_{\pm}R_{n \pm 1,\ell}~(r),
\label{m32}
\end{equation}
where
\begin{equation}
\ell_{-}=  \sqrt{\frac{ n(n + \beta_{\ell})( n + 2 \beta_{\ell} + 1)}{( n +  \beta_{\ell} + 1)}},~~~~ \ell_{+}= \sqrt{\frac{ (n + 1)(n + \beta_{\ell} + 2)( n + 2 \beta_{\ell} + 2)}{( n +  \beta_{\ell} + 1)}}~.
\label{m33}
\end{equation}

On studying the dynamical group associated to the annihilation and creation operators $\hat{\cal{L}}_{-}$ and $\hat{\cal{L}}_{+}$ and based on the results of equations (\ref{m32}) and (\ref{m33}), we can evaluate the commutator $[\hat{\cal{L}}_{-},~\hat{\cal{L}}_{+}]$ as:\\
\begin{equation}
[\hat{\cal{L}}_{-},~\hat{\cal{L}}_{+} ]R_{n, \ell}~(r)
=2 \ell_{0}R_{n, \ell}~(r),
\label{m34}
\end{equation}
where
\begin{equation}
\ell_{0}= (n + \beta_{\ell} + 1 )
\label{m35}
\end{equation}
and $\hat{\cal{L}}_{0}$ is defined as
\begin{equation}
\hat{\cal{L}}_{0}= (\hat{n} + \beta_{\ell} + 1).
\label{m36}
\end{equation}

Thus, operators $\hat{\cal{L}}_{\mp}$ and $\hat{\cal{L}}_{0}$ satisfy the following commutation relations:
\begin{equation}
\displaystyle{[\hat{\cal{L}}_{0},~\hat{\cal{L}}_{\mp} ]R_{n, \ell}~(r)
=\mp\hat{\cal{L}}_{\mp} R_{n \mp 1, \ell}~(r)}.
\label{m37}
\end{equation}
The action of $\hat{\cal{L}}_{+}$ on the radial wavefunction $R_{n, \ell}~(r)$ gives
\begin{equation}
\begin{array}{rl}
&\left( \hat{\cal{L}}_{0} \hat{\cal{L}}_{+} - \hat{\cal{L}}_{+}\hat{\cal{L}}_{0} \right)R_{n, \ell}~(r) = \hat{\cal{L}}_{+} R_{n, \ell}~(r)\\
& = \hat{\cal{L}}_{0} \left(\hat{\cal{L}}_{+}R_{n, \ell}~(r) \right) - \ell_{0} \left(\hat{\cal{L}}_{+}R_{n, \ell}~(r)\right)\\
&= (\ell_{0} + 1)\hat{\cal{L}}_{+} R_{n, \ell}~(r)
\label{m38}.
\end{array}
\end{equation}
If $\hat{\cal{L}}_{+} R_{n, ~\ell}~(r)$ is non-zero, then, it is an eigenfunction of $\hat{\cal{L}}_{0}$ with eigenvalue $(\ell_{0} + 1)$. Thus, the effect of $\hat{\cal{L}}_{+}$ is to raise the eigenvalue by one unit.
Similarly, the action of $\hat{\cal{L}}_{-}$ on the radial wavefunction $R_{n, \ell}~(r)$ gives
\begin{equation}
\begin{array}{rl}
&\left( \hat{\cal{L}}_{0} \hat{\cal{L}}_{-} - \hat{\cal{L}}_{-}\hat{\cal{L}}_{0} \right)R_{n, \ell}~(r) = - \hat{\cal{L}}_{-}~ R_{n, \ell}~(r)\\
& = \hat{\cal{L}}_{0} \left(\hat{\cal{L}}_{-}~R_{n, \ell}~(r) \right) - \ell_{0} \left(\hat{\cal{L}}_{-}~R_{n, \ell}~(r)\right)\\
&= (\ell_{0} - 1)\hat{\cal{L}}_{-}~ R_{n, \ell}~(r)
\label{m39},
\end{array}
\end{equation}
and if $\hat{\cal{L}}_{+} R_{n, ~\ell}~(r)$ is non-zero, then, it is an eigenfunction of $\hat{\cal{L}}_{0}$ with eigenvalue $(\ell_{0} - 1)$ ~(this is the reason for calling $\hat{\cal{L}}_{+}$ and $\hat{\cal{L}}_{-}$ raising and lowering operators respectively).

For the Hermitian operators, we define the operators as follows:
\begin{equation}
\displaystyle{\hat{\cal{L}}_{x} = \frac{1}{2} \left( \hat{\cal{L}}_{+} + \hat{\cal{L}}_{-} \right)},~~~
\displaystyle{\hat{\cal{L}}_{y} = \frac{1}{2 i} \left( \hat{\cal{L}}_{+} - \hat{\cal{L}}_{-} \right)},~~~
\displaystyle{\hat{\cal{L}}_{z} = \hat{\cal{L}}_{0}}
\label{m40},
\end{equation}
we obtained the following commutation relations
\begin{equation}
\displaystyle{[\hat{\cal{L}}_{x},~\hat{\cal{L}}_{y} ] = - i \hat{\cal{L}}_{z}},~~
\displaystyle{[\hat{\cal{L}}_{y},~\hat{\cal{L}}_{z} ] = i \hat{\cal{L}}_{x}},~~
\displaystyle{[\hat{\cal{L}}_{z},~\hat{\cal{L}}_{x} ] = i \hat{\cal{L}}_{y}}.
\label{m41}
\end{equation}

The Casimir operator  \cite{Cas31} can be expressed as
\begin{equation}
\begin{array}{rl}
&\hat{\cal{C}}R_{n, \ell}~(r)
=\left( \hat{\cal{L}}_{0}(\hat{\cal{L}}_{0} - 1 ) - \hat{\cal{L}}_{+} \hat{\cal{L}}_{-}\right) R_{n, \ell}~(r)
\\ \nonumber
&=\left(\hat{\cal{L}}_{0}( \hat{\cal{L}}_{0} + 1) - \hat{\cal{L}}_{-} \hat{\cal{L}}_{+} \right)R_{n, \ell}~(r) \\ \nonumber
&=\beta_{\ell}~(\beta_{\ell} + 1)R_{n, \ell}~(r).
\label{m42}
\end{array}
\end{equation}
The Casimir operator $\hat{\cal{C}} $ now satisfies
\begin{equation}
[ \hat{\cal{C}},  \hat{\cal{L}}_{\pm}] =[ \hat{\cal{C}},  \hat{\cal{L}}_{x}] =
[\hat{\cal{C}}, \hat{\cal{L}}_{y}] =  [\hat{\cal{C}}, \hat{\cal{L}}_{z}] = 0,
\label{m43}
\end{equation}
the operators $\hat{\cal{L}}_{\pm} $, $\hat{\cal{L}}_{x} $, $\hat{\cal{L}}_{y} $, $\hat{\cal{L}}_{z} $ and $\hat{\cal{L}}_{0} $ satisfy the commutation relations of the dynamical group $SU(1,1)$ algebra, which is isomorphic to an $SO(2, 1)$ algebra (i. e. $SU(1,1)$ $\sim$ $SO(2, 1)$. The commutation rules are valid for the infinitesimal operators of the non-compact group $SU(1, 1)$ \cite{BaR00, NiE91}.

These relations coincide with the formulas that define the action of the infinitesimal operators $\hat{\cal{L}}_{\pm} $ and $\hat{\cal{L}}_{0} $ of the dynamical group $SU(1, 1)$ on a basis $\mid j, k \rangle$ of the irreducible representation $D^{+}(j)$  belonging to the discrete positive series in an abstract
 Hilbert space \cite{Don07,BaR00,Ada94,NiE91}.  The eigenvalues have the ground state and therefore, the representation of the dynamical group $SU(1, 1)$ belongs to $D^{+}(j)$, using the Dirac notation $\mid j, k\rangle = R_{n, \ell}~(r)$ :
\begin{equation}
\begin{array}{rl}
&\displaystyle{\hat{\cal{C}}\mid j, k\rangle = j(j + 1)\mid j, k\rangle }
\\ \nonumber
&\displaystyle{\hat{\cal{L}}_{0}\mid j, k\rangle = \ell_{0}\mid j, k\rangle = k \mid j, k\rangle} \\ \nonumber
&\displaystyle{\hat{\cal{L}}_{\pm}\mid j, k\rangle = \left[\frac{k(k \pm 1)^{2} - j(j + 1)(k \pm 1) }{k} \right]^{\frac{1}{2}}  \mid j, k \pm 1 \rangle} \\ \nonumber
&k= - j + n, ~~n= 1, 2, ~~\ldots, j<0.
\label{m44}
\end{array}
\end{equation}

Furthermore, the following expressions can be easily obtained from the operators $\hat{\cal{L}}_{\mp}$ and $\hat{\cal{L}}_{0}$ as follows:
\begin{equation}
\begin{array}{rl}
&\displaystyle{r= \frac{1}{ \xi}[2 \hat{\cal{L}}_{0} -(\hat{\cal{L}}_{+} + \hat{\cal{L}}_{-})]}
\\ \nonumber
&\displaystyle{r \frac{d}{dr}=  \frac{1}{2}(\hat{\cal{L}}_{+} - \hat{\cal{L}}_{-}) - 1}.
\label{m45}
\end{array}
\end{equation}
With these, the matrix elements for $r$ and $r\frac{d}{dr}$ are obtained as follows:
\begin{equation}
\langle R_{m, \ell}~(r) \mid r\mid R_{n, \ell}~(r)\rangle
=\displaystyle{ \frac{1}{\xi}  \left[  ( n + \beta_{\ell} + 1) \delta_{m, n} - \ell_{+} \delta_{m, n + 1}  - \ell_{-} \delta_{m, n - 1} \right]}
\label{m46}
\end{equation}
and
\begin{equation}
\langle R_{m, \ell}~(r) \mid r\frac{d}{dr}\mid R_{n, \ell}~(r)\rangle
 =\displaystyle{ \frac{\ell_{+}}{2}~ \delta_{m, n + 1} - \frac{\ell_{ -}}{2}~ \delta_{m, n - 1}  -  \delta_{m, n}}.
\label{m47}
\end{equation}
From equations (\ref{m46}) and (\ref{m47}), We can deduce the following relations:
\begin{equation}
\begin{array}{rl}
&\displaystyle{\xi \langle R_{m, \ell}~(r) \mid r\mid R_{n, \ell}~(r)\rangle + \langle R_{m, \ell}~(r) \mid r\frac{d}{dr}\mid R_{n, \ell}~(r)\rangle} \nonumber \\
&=\displaystyle{( n +  \beta_{\ell}~ ) \delta_{m, n} - \frac{1}{2} \ell_{+} \delta_{m, n + 1} - \frac{3 \ell_{-}}{2} \delta_{m, n - 1}}
\label{m48}
\end{array}
\end{equation}
and
\begin{equation}
\begin{array}{rl}
&\displaystyle{\xi \langle R_{m, \ell}~(r) \mid r\mid R_{n, \ell}~(r)\rangle - \langle R_{m, \ell}~(r) \mid r\frac{d}{dr}\mid R_{n, \ell}~(r)\rangle} \nonumber \\
&=\displaystyle{( n +  \beta_{\ell} + 1 ) \delta_{m, n} - \frac{3}{2} \ell_{+} \delta_{m, n + 1} - \frac{ \ell_{-}}{2} \delta_{m, n - 1}},
\label{m49}
\end{array}
\end{equation}
these relations form a useful link for finding the matrix elements from ladder operators.

\section{Numerical calculations of the energy levels and matrix elements.}

\subsection{Energy eigenvalues of the Kratzer and modified Kratzer potentials.}

The energy eigenvalues for the Kratzer and modified Kratzer potentials are obtained respectively, as :
\begin{equation}
E_{n, \ell}^{K}  = -~ \frac{ 2 \mu D_{0}^{2}r_{0}^{2}}{ \hbar^{2}  \left[ n + \beta_{\ell} + 1 \right]^{2}}
\label{m50}
\end{equation}
and
\begin{equation}
E_{n, \ell}^{MK}  =  -~\frac{ 2 \mu D_{0}^{2}r_{0}^{2}}{ 2 \hbar^{2}  \left[ n + \beta_{\ell} + 1 \right]^{2}} + D_{0},
\label{m51}
\end{equation}
where
\begin{equation}
\beta_{\ell}  =  \frac{1}{2} \left[-1 + \sqrt{( 2 \ell + 1)^{2}  + \frac{8 \mu D_{0}r_{0}^{2}}{\hbar^{2}}}~~  \right].
\label{m52}
\end{equation}

\begin{table}[!hbp]
\caption{Reduced masses and spectroscopic properties of the $CO$ and $NO$
diatomic molecules in the ground electronic state.
The data listed in this table are taken from \cite{KaP70,BrR69}.\vspace*{13pt}} {\small
\begin{tabular}{|c|c|c|}
\hline
{}&{}   &{}\\[-1.5ex]
Parameters & $CO$ & $NO$ \\[1ex]
\hline
$D_{o}$(in eV)     &10.84514471    &8.043782568   \\[1ex]
$r_{o}$(in $A^{o}$)&1.1282         &1.1508        \\[1ex]
$\mu$(in amu)      &6.860586000    &7.468441000   \\[1ex]\hline
\end{tabular}\label{tab1} }
\vspace*{-13pt}
\end{table}

In this work, energy eigenvalues for $CO$ and $NO$ diatomic
molecules for the various values of $n$ and $\ell$ are obtained by
means of the factorization method (FM) (equations (50) and (51) ) with
the parameters given in Table 1. The results obtained are compared
with other results obtained by using: AIM method \cite{BaE07};
Nikiforov-Uvarov (NU) method \cite{BeE06} and Exact quantization
rule (EQR) method \cite{IkS08}.

Table $2$ shows the exact bound state energy eigenvalues of the
Kratzer potential for $CO$ and $NO$ diatomic molecules for various
values of $n$ and $\ell$ using equation (50) obtained by
factorization method (FM) and other results obtained by using AIM
\cite{BaE07} and EQR \cite{IkS08} methods. Similarly, Table $3$
shows the exact bound state energy eigenvalues of the modified
Kratzer potential for $CO$ and $NO$ diatomic molecules for various
values of $n$ and $\ell$ using equation (51)  obtained by
factorization method (FM) and other results obtained by using AIM
\cite{BaE07} and NU \cite{BeE06} methods.

\begin{table}[!hbp]
\caption{Comparison of the energy eigenvalues (in eV), corresponding to the Kratzer potential for various $n$ and $\ell$ quantum numbers for $CO$ and $NO$ diatomic molcules, where $\hbar c=1973.29eV A^{o}$.\vspace*{13pt}} {\scriptsize
\begin{tabular}{|c|c|c|c|c|c|c|c|}
\hline
{}&{} &{} &{}&{}&{}&{} &{}\\[-1.5ex]
$n$ & $\ell$ & $CO$ [FM]   & $CO$ [EQR] &$CO$[AIM] &$NO$[FM]&$NO$[EQR]&$NO$[AIM]\\[1ex]
\hline
$0$& $0$ &-10.79431534387622 &-10.794315323 &-10.79431532 &-8.002658755212952& -8.002659419493& -8.002659417 \\[1ex]
$1$     & $0$  &-10.69383913769446     &-10.693839925   &-10.69383992
& -7.921456003136883     &-7.921456840689         &-7.921456839      \\[1ex]
$1$     & $1$  &-10.69337109882925     &-10.693371229   &-10.69337123
&-7.921042972272428      &-7.921043829925         &-7.921043834  \\[1ex]
$2$     & $0$  &-10.59476059512928     &-10.594760890   &-10.59476089
&-7.841483226715529      &-7.841483958093         &-7.841483956  \\[1ex]
$2$     & $1$  &-10.59429734443464     &-10.594298692   &-10.59429869
&-7.841075958492119      &-7.841077185904         &-7.841077188  \\[1ex]
$2$     & $2$  &-10.59337288634942     &-10.593374417   &-10.59337441
&-7.840262393682968      &-7.840263768523         &-7.840263771  \\[1ex]
$3$     & $0$  &-10.49705105930509     &-10.497052462   &-10.49705246
&-7.76271486692960      &-7.762716067159         &-7.762716066  \\[1ex]
$3$     & $1$  &-10.49659537021400     &-10.496596643   &-10.49659664
& -7.762314093207729     &-7.762315408528         &-7.762315413  \\[1ex]
$3$     & $2$  &-10.49568316746262     &-10.495685124   &-10.49568512
&-7.761512650377457      & -7.761514215884        &-7.761514218  \\[1ex]
$3$     & $3$  &-10.49431563207918     &-10.494318144   &-10.49431814
&-7.760311623981464      &-7.760312738370         &-7.760312744  \\[1ex]
$4$     & $0$  &-10.40068823210378     &-10.400689478   &-10.40068947
&-7.685127711083608      & -7.685129080626        &-7.685129079  \\[1ex]
$4$     & $1$  &-10.40023809235447     &-10.400239921   &-10.40023992
&-7.684732464238659      & -7.684734413653        &-7.684734417  \\[1ex]
$4$     & $2$  &-10.39933885575065     &-10.399340924   &-10.39934092
&-7.683942903888297      & -7.683945202003        &-7.683945203  \\[1ex]
$4$     & $3$  &-10.39798982948594     &-10.397992722   &-10.39799272
&-7.682760084230797      & -7.682761690175        &-7.682761696  \\[1ex]
$4$     & $4$  &-10.39619321456787     &-10.396195666   &-10.39619567
&-7.681181936089156      & -7.681184244677        &-7.681184246  \\[1ex]
$5$     & $0$  &-10.30564563709163     &-10.305647347   &-10.30564735
&-7.608697429711712      &-7.608699510108         &-7.608699509  \\[1ex]
$5$     & $1$  &-10.30520186777971     &-10.305203938   &-10.30520394
&-7.608308414224594      & -7.608310715917        &-7.608310719  \\[1ex]
$5$     & $2$  &-10.30431444380035     &-10.304317236   &-10.30431723
&-7.607531297904489      &-7.607533247563         &-7.607533248  \\[1ex]
$5$     & $3$  &-10.30298450330134     &-10.302987469   &-10.30298747
&-7.606365523492213      &-7.606367345012         &-7.606367349  \\[1ex]
$5$     & $4$  &-10.30121238898166     &-10.301214985   &-10.30121499
&-7.604810653591787      &-7.604813367976         &-7.604813368  \\[1ex]
$5$     & $5$  &-10.29899764904374     &-10.299000242   &-10.29900024
&-7.602869548627632      &-7.602871795644         &-7.602871795  \\[1ex]\hline
\end{tabular}\label{tab2} }
\vspace*{-13pt}
\end{table}

\begin{table}[!hbp]
\caption{Comparison of the energy eigenvalues (in eV), corresponding to the modified Kratzer potential for various $n$ and $\ell$ quantum numbers for $CO$ and $NO$ diatomic molcules, where $\hbar c=1973.29eV A^{o}$.\vspace*{13pt}} {\scriptsize
\begin{tabular}{|c|c|c|c|c|c|c|c|}
\hline
{}&{} &{} &{}&{}&{}&{} &{}\\[-1.5ex]
$n$ & $\ell$ & $CO$ [FM]   & $CO$ [EQR] &$CO$[NU] &$NO$[FM]&$NO$[EQR]&$NO$[NU]\\[1ex]
\hline
$0$&$0$
&0.05082892797436500      &0.050829386733     &0.050823
&0.04112347897894253      &0.041123148507   &0.041118 \\[1ex]
$1$&$0$
&0.1513051341561269     &0.151304784801     &0.151287
&0.1223262310550117      &0.122325727312   &0.122311      \\[1ex]
$1$&$1$
&0.1517731730213381    &0.151773481462     &0.151755
&0.1227392619194667      &0.122738738076   &0.122724  \\[1ex]
$2$&$0$
&0.2503836767213095     &0.250383819984     &0.250354
&0.2022990074763653     &0.202298609907   &0.202274  \\[1ex]
$2$&$1$
&0.2508469274159442     &0.250846018075     &0.250816
&0.2027062756997760      &0.202705382097   &0.202681  \\[1ex]
$2$&$2$
&0.2517713855011650      &0.251770292931     &0.251744
&0.2035198405089265     &0.203518799478   &0.203494  \\[1ex]
$3$&$0$
&0.3480932125454999      &0.348092247640     &0.348051
&0.2810673574989346      &0.281066500842   &0.281033  \\[1ex]
$3$&$1$
&0.3485489016365868      &0.348548066746     &0.348507
&0.2814681409841651     &0.281467159473   &0.281434  \\[1ex]
$3$&$2$
&0.3494611043879683      &0.349459585720     &0.349418
&0.2822695838144371      &0.282268352117  &0.282235  \\[1ex]
$3$&$3$
&0.3508286397714091     &0.350826566166     &0.350785
&0.2834706102104301     &0.283469829631   &0.283436  \\[1ex]
$4$&$0$
&0.4444560397468020      &0.444455232045     &0.444403
&0.3586545231082869      & 0.358653487375  &0.358611  \\[1ex]
$4$&$1$
&0.4449061794961153      &0.444904789014     &0.444852
&0.3590497699532351     &0.359048154348  &0.359006  \\[1ex]
$4$&$2$
&0.4458054160999403     &0.445803785755     &0.445751
&0.3598393303035978      &0.359837365998  &0.359795  \\[1ex]
$4$&$3$
&0.4471544423646421      &0.447151987956     &0.447099
&0.3610221499610970      &0.361020877826  &0.360978  \\[1ex]
$4$&$4$
&0.4489510572827147      &0.448949044337     &0.448895
&0.3626002981027385     & 0.362598323324  &0.362555  \\[1ex]
$5$&$0$
&0.5394986347589601     &0.539497362596     &0.539434
&0.4350848044801827      &0.435083057893   &0.435032  \\[1ex]
$5$&$1$
&0.5399424040708798     &0.539940771611     &0.539877
&0.4354738199673003     &0.435471852084  &0.435421  \\[1ex]
$5$&$2$
&0.5408298280502333      &0.540827474443     &0.540764
&0.4362509362874052     &0.436249320438   &0.436198  \\[1ex]
$5$&$3$
&0.5421597685492472     &0.542157240772     &0.542093
&0.4374167106996811     &0.437415222989   &0.437364  \\[1ex]
$5$&$4$
&0.5439318828689306      &0.543929725307     &0.543865
&0.4389715806001080      &0.438969200025   &0.438917  \\[1ex]
$5$&$5$
&0.5461466228068463     &0.546144468004    &0.546082
&0.4409126855642622      &0.440910772357  &0.440858  \\[1ex]\hline
\end{tabular}\label{tab3} }
\vspace*{-13pt}
\end{table}

\subsection{Matrix elements of $r$ and $r\frac{d}{dr}$ for the two potentials.}
The matrix elements for $r$ and $r\frac{d}{dr}$ are given as:
\begin{equation}
\begin{array}{rl}
&\displaystyle{\langle R_{n, \ell}~(r) \mid r\mid R_{n, \ell}~(r)\rangle}   \nonumber \\
&=
\displaystyle{ \frac{1}{\xi}  \left[  ( n + \beta_{\ell} + 1) - \sqrt{\frac{ (n + 2)(n + \beta_{\ell} + 3)( n + 2 \beta_{\ell} + 3)}{( n +  \beta_{\ell} + 2~)}}  - \sqrt{\frac{ (n - 1)(n + \beta_{\ell} -1)( n + 2 \beta_{\ell}~)}{( n +  \beta_{\ell}~ )}} \right]}
\label{m53}
\end{array}
\end{equation}
and
\begin{equation}
\begin{array}{rl}
&\displaystyle{\langle R_{n, \ell}~(r) \mid r\frac{d}{dr}\mid R_{n, \ell}~(r)\rangle} \nonumber \\
& = \displaystyle{\frac{1}{2}\sqrt{\frac{ (n + 2)(n + \beta_{\ell} + 3)( n + 2 \beta_{\ell} + 3)}{( n +  \beta_{\ell} + 2)}} - \frac{1}{2}\sqrt{\frac{ (n - 1)(n + \beta_{\ell} -1)( n + 2 \beta_{\ell}~)}{( n +  \beta_{\ell}~ )}}  - 1},
\label{m54}
\end{array}
\end{equation}
where
\begin{equation}
\xi= -~ \frac{4 \mu D_{0}r_{0} }{\hbar^{2}(n + \beta_{\ell} + 1)} ~.
\label{m55}
\end{equation}
With equations (\ref{m53}) and (\ref{m54}), we can deduce the following relations:
\begin{equation}
\begin{array}{rl}
&\displaystyle{ \Gamma_{1}= \xi \langle R_{n, \ell}~(r) \mid r\mid R_{n, \ell}~(r)\rangle + \langle R_{n, \ell}~(r) \mid r\frac{d}{dr}\mid R_{n, \ell}~(r)\rangle} \nonumber \\
&=\displaystyle{( n +  \beta_{\ell}~ ) - \frac{1}{2} \sqrt{\frac{ (n + 2)(n + \beta_{\ell} + 3)( n + 2 \beta_{\ell} + 3)}{( n +  \beta_{\ell} + 2)}} - \frac{3}{2} \sqrt{\frac{ (n - 1)(n + \beta_{\ell} -1)( n + 2 \beta_{\ell}~)}{( n +  \beta_{\ell}~ )}}}
\label{m56}
\end{array}
\end{equation}
and
\begin{equation}
\begin{array}{rl}
&\displaystyle{\Gamma_{2}= \xi \langle R_{n, \ell}~(r) \mid r\mid R_{n, \ell}~(r)\rangle - \langle R_{n, \ell}~(r) \mid r\frac{d}{dr}\mid R_{n, \ell}~(r)\rangle} \nonumber \\
&=\displaystyle{( n +  \beta_{\ell}~ + 1) - \frac{1}{2} \sqrt{\frac{ (n + 2)(n + \beta_{\ell} + 3)( n + 2 \beta_{\ell} + 3)}{( n +  \beta_{\ell} + 2)}} - \frac{3}{2} \sqrt{\frac{ (n - 1)(n + \beta_{\ell} -1)( n + 2 \beta_{\ell}~)}{( n +  \beta_{\ell}~ )}}}.
\label{m57}
\end{array}
\end{equation}
In this case, the matrix elements for the two potentials (Kratzer and modified Kratzer potentials) give the same results, since these matrix elements depend on $b = D_{0}r_{0}$ and $a = - 2D_{0}r_{0}$ only, and not on $c = D_{0}$. Hence, when the Kratzer potential is shifted by the amount $D_{0}$, (that is, the modified Krazter potential), it has no effect on the matrix elements. For the numerical results, see Tables $4$ and $5$.

\newpage
\begin{table}[!hbp]
\caption{The values of the $\langle R_{n, ~\ell}\mid r \mid R_{n, ~\ell}\rangle$ [Equation (53)], $\langle R_{n, ~\ell}\mid r\frac{d}{dr} \mid R_{n, ~\ell}\rangle$ [Equation (54)], $\Gamma_{1}$ [Equation (56)] and $\Gamma_{2}$  [Equation (57)], corresponding to the Kratzer and modified Kratzer potentials for various $n$ and $\ell$ quantum numbers for $CO$ diatomic molcule, where $\hbar c=1973.29eV A^{o}$.\vspace*{13pt}} {\scriptsize
\begin{tabular}{|c|c|c|c|c|c|}
\hline
{}&{} &{} &{}&{}  &{} \\[-1.5ex]
$n$ & $\ell$ & $\langle R_{n, ~\ell}\mid r \mid R_{n, ~\ell}\rangle$  & $\langle R_{n,~ \ell}\mid r\frac{d}{dr} \mid R_{n, ~\ell}\rangle$  & $\Gamma_{1} $  & $\Gamma_{2}$ \\[1ex] \hline

$1$ & $0$ &-0.4761490924054464 &16.97323506451677 &195.3895273866551 &161.4430572576216 \\[1ex]
$1$ & $1$ &-0.4761710273791221 &16.97343118040934 &195.3940309777938&161.4471686169751 \\[1ex]
$2$ & $0$ &-0.4107558210705384 &9.473123928272610 &162.6714457007481&143.7251978442029
\\[1ex]
$2$ & $1$ &-0.4107755614421622 &9.473236107448747 &162.6755771602085&143.7291049453111
\\[1ex]
$2$ & $2$ &-0.4108150436563324 &9.473460462197114 &162.6838400978704&143.7369191734761
\\[1ex]
$3$ & $0$ &-0.3789129917968390 &7.666855095230551 &148.3356842660929&133.0019740756318
\\[1ex]
$3$ & $1$ &-0.3789316534237137 &7.666946649417545 &148.3396482090482&133.0057549102131
\\[1ex]
$3$ & $2$ &-0.3789689781099853 &7.667129754891770 &148.3475761191884&133.0133166094049
\\[1ex]
$3$ & $3$ &-0.3790249687207100 &7.667404405854176 &148.3594680449690&133.0246592332607
\\[1ex]
$4$ & $0$ &-0.3533372017343955 &6.615793772861618 &137.1863080576614&123.9547205119381
\\[1ex]
$4$ & $1$ &-0.3533549972979473 &6.615873160774083 &137.1901403213880&123.9583939998398
\\[1ex]
$4$ & $2$ &-0.3533905898264022 &6.616031934119029 &137.1978048773512&123.9657410091131
\\[1ex]
$4$ & $3$ &-0.3534439821224761 &6.616270087936851 &137.2093017825661&123.9767616066924
\\[1ex]
$4$ & $4$ &-0.3535151205980674 &6.616587357061658 &137.2246186798360&123.9914439657127
\\[1ex]
$5$ & $0$ &-0.3312181486228352 &5.899103372782779 &127.7353299114129&115.9371231658474
\\[1ex]
$5$ & $1$ &-0.3312351970446675 &5.899174359294310 &127.7390497423004&115.9407010237118
\\[1ex]
$5$ & $2$ &-0.3312692952626132 &5.899316330129842 &127.7464894362121&115.9478567759524
\\[1ex]
$5$ & $3$ &-0.3313204460252490 &5.899529280914656 &127.7576490574162&115.9585904955868
\\[1ex]
$5$ & $4$ &-0.3313885980893305 &5.899812974634305 &127.7725166245981&115.9728906753295
\\[1ex]
$5$ & $5$ &-0.3314738123034079 &5.900167633018892 &127.7911043436078&115.9907690775700
\\[1ex]\hline
\end{tabular}\label{tab4} }
\vspace*{-13pt}
\end{table}

\newpage
\begin{table}[!hbp]
\caption{The values of the $\langle R_{n, ~\ell}\mid r \mid R_{n, ~\ell}\rangle$ [Equation (53)], $\langle R_{n, ~\ell}\mid r\frac{d}{dr} \mid R_{n, ~\ell}\rangle$ [Equation (54)], $\Gamma_{1}$ [Equation (56)] and $\Gamma_{2}$  [Equation (57)], corresponding to the Kratzer and modified Kratzer potentials for various $n$ and $\ell$ quantum numbers for $NO$ diatomic molcule, where $\hbar c=1973.29eV A^{o}$.\vspace*{13pt}} {\scriptsize
\begin{tabular}{|c|c|c|c|c|c|}
\hline
{}&{} &{} &{}&{}  &{} \\[-1.5ex]
$n$ & $\ell$ & $\langle R_{n, ~\ell}\mid r \mid R_{n, ~\ell}\rangle$  & $\langle R_{n, ~\ell}\mid r\frac{d}{dr} \mid R_{n, ~\ell}\rangle$  & $\Gamma_{1} $  & $\Gamma_{2}$ \\[1ex] \hline

$1$ & $0$ &-0.4819553974463722 &16.21611432627930 &178.3853413621973 &145.9531127096387 \\[1ex]
$1$ & $1$ &-0.4819818755489971 &16.21633768030118 &178.3902449613004 &145.9575696006980
\\[1ex]
$2$ & $0$ &-0.4124035377598499 &9.040224937878095 &147.1042625831576 &129.0238127074014
\\[1ex]
$2$ & $1$ &-0.4124272376527343 &9.040352590403065 &147.1087421492765 &129.0280369684704
\\[1ex]
$2$ & $2$ &-0.4124746395526131 &9.040607890585319 &147.1177013069172 &129.0364855257466
\\[1ex]
$3$ & $0$ &-0.3785773170199991 &7.313655510582683 &133.4152241892145 &118.7879131680491
\\[1ex]
$3$ & $1$ &-0.3785996534470747 &7.313759628367830 &133.4195129653269 &118.7919937085912
\\[1ex]
$3$ & $2$ &-0.3786443283574571 &7.313967860029029 &133.4280905503876 &118.8001548303296
\\[1ex]
$3$ & $3$ &-0.3787112793749515 &7.314279887895779 &133.4409442456685 &118.8123844698770
\\[1ex]
$4$ & $0$ &-0.3514058390444385 &6.309618628486895 &122.7741267465619 &110.1548894895881
\\[1ex]
$4$ & $1$ &-0.3514270813879861 &6.309708856787779 &122.7782656025786 &110.1588478890030
\\[1ex]
$4$ & $2$ &-0.3514695680843574 &6.309889310052519 &122.7865433532380 &110.1667647331330
\\[1ex]
$4$ & $3$ &-0.3515332399199089 &6.310159713089281 &122.7989477575504 &110.1786283313719
\\[1ex]
$4$ & $4$ &-0.3516182293790348 &6.310520592939504 &122.8155035676796 &110.1944623818006
\\[1ex]
$5$ & $0$ &-0.3278978812775772 &5.625407532078025 &113.7570644421939 &102.5062493780379
\\[1ex]
$5$ & $1$ &-0.3279181800627655 &5.625488165029850 &113.7610753342060 &102.5100990041463
\\[1ex]
$5$ & $2$ &-0.3279587796016631 &5.625649427995388 &113.7690971617573 &102.5177983057665
\\[1ex]
$5$ & $3$ &-0.3280196234075503 &5.625891075135918 &113.7811180744773 &102.5293359242055
\\[1ex]
$5$ & $4$ &-0.3281008382258643 &5.626213577576859 &113.7971620775729 &102.5447349224192
\\[1ex]
$5$ & $5$ &-0.3282023110716663 &5.626616443662535 &113.8172054699764 &102.5639725826513
\\[1ex]\hline
\end{tabular}\label{tab5} }
\vspace*{-13pt}
\end{table}

\section{Conclusions}

In this paper, I have studied the eigenvalues, the eigenfunctions and the matrix elements of the Kratzer-type molecular potentials (generalized Kratzer molecular potentials). The ladder (creation and annihilation) operators for the radial wavefunctions are established. Also, the Hermitian operators of these ladder operators are obtained. These operators satisfy the commutation relations of an $SU(1, 1)$ dynamical group, the action of $\hat{\cal{L}}_{\pm}$ on the wavefunctions that reveals the fact about the raising and lowering effect are established. The matrix elements of the different functions $r$ and $r \frac{d}{dr}$ are also obtained from the ladder operators in a closed form.

The solutions of the generalized Kratzer molecular potentials are obtained via an $SU(1, 1)$ algebraic approach and the results can be generalized to other form of potentials in equation (\ref{m5}). The generalized Kratzer potential model proposed in this work \cite{IkS071,IkS072}, allows one to obtain the eigenvalues, the matrix elements and the radial eigenfunctions for the two important molecular potentials simultaneously (that is, the Kratzer and modified Kratzer potentials).

This generalization covers the descriptions about the two potentials, the shapes of the two potentials for $CO$ and $NO$ diatomic molecules are also shown in figure $1$.  More importantly, the generalized Kratzer potential model used in this work generates results for the two molecular potentials. In addition, the numerical results are obtained for the eigenvalues for $CO$ and $NO$ diatomic molecules for the two molecular potentials. The results obtained are compared with other existing results (EQR, AIM and NU), and the results agree favourably with other results, see Tables $2$ and $3$.

Tables $4$ and $5$ show the matrix elements of $r$, $r\frac{d}{dr}$, their sum ($\Gamma_{1}$) and their difference ($\Gamma_{2}$) obtained for $CO$ and $NO$ diatomic molecules, the results obtained are the same results for the two molecular potentials. This is because, the matrix elements depend on $b = D_{0}r_{0}$ and $a = - 2D_{0}r_{0}$ only, and not on $c = D_{0}$. Hence, the shifted amount $D_{0}$, (in the case of the modified Krazter potential) has no effect on the matrix elements.

The advantage of the present approach is that it enables one to find the energy eigenvalues, eigenfunctions and matrix elements in a simple way. The approach presented in this  study is efficient and is a very useful link for finding the matrix elements from ladder operators. This approach can be used to find the energy eigenvalues, eigenfunctions and radial matrix elements of the Schr\"{o}dinger equation with a given exactly solvable molecular potentials for various diatomic molecules for any values of $n$ and $\ell$ quantum numbers.

The ladder operators constructed in this study, are very useful tools in quantum-mechanical calculations of the various matrix elements based on the Kratzer molecular basis function. Furthermore, these operators can be used in constructing coherent states.
\vspace{5mm}

[Note to the desk Editor: Please, put the figure in the section $2$,
where it is indicated in the text]

\vspace{5mm}
{\bf  \huge{Acknowledgements}.}

I am very grateful to the ICTP for my postdoctoral position at the ICMPA-UNESCO chair
under the Prj -$15$ ICTP project, this work is a product of the motivations I received
from Prof. M. N. Hounkonnou (The Chair, ICMPA, Universite d'Abomey-Calavi, Benin) and
other ICMPA-UNESCO chair visiting Professors. I appreciate the efforts of
 Profs. Dong, S. H.; Sen, K. D.; L$\grave{o}$pez-Bonila, J. L. and Emeritus Prof. K. T. Hecht
 for sending valuable materials to me. I thank the anonymous referees for the careful
 reading of the manuscript and the suggestions that improved the paper.


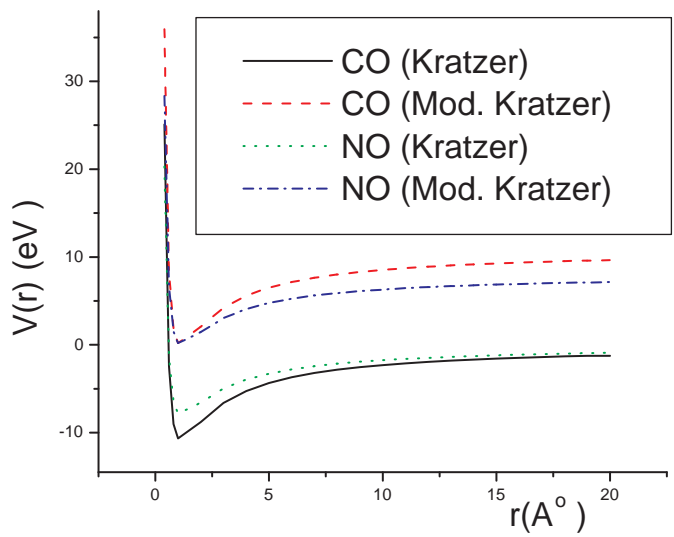
\end{document}